\documentclass[a4paper, 10 pt, conference]{ieeeconf}  % Comment this line out if you need a4paper
\IEEEoverridecommandlockouts % This command is only needed if you want to use the \thanks command
\usepackage[polish,american]{babel}
\usepackage{graphicx}
\usepackage{array}
\usepackage{tablefootnote}
\usepackage{footmisc}

\usepackage{mathtools}
\usepackage{amsmath}
\usepackage{amsfonts}
\usepackage{braket}

\usepackage[hidelinks]{hyperref}
\usepackage{cleveref}
\crefname{equation}{}{}
\crefname{figure}{Fig.}{Fig.}
\crefname{table}{Tab.}{Tab.}
\crefname{section}{Section}{Sections}

\usepackage{orcidlink}

\usepackage{color}

\newcommand{\ceil}[1]{\lceil#1\rceil}

\title{\LARGE \bf
    FRQI Pairs method for image classification \\ 
    using Quantum Recurrent Neural Network
}

\author{
    Rafał Potempa\orcidlink{0000-0002-0813-0606}$^{1,2}$,
    Michał Kordasz\orcidlink{0000-0003-4193-9244}$^{1}$,
    Sundas Naqeeb Khan\orcidlink{0000-0002-6973-1843}$^{1}$,
    Krzysztof Werner\orcidlink{0000-0002-0327-837X}$^{1}$, \\
    Kamil Wereszczyński\orcidlink{0000-0003-1686-472X}$^{1}$,
    Krzysztof Simiński\orcidlink{0000-0002-6118-606X}$^{1}$,
    Krzysztof A. Cyran\orcidlink{0000-0003-1789-4939}$^{1}$% <-this % stops a space
    \thanks{$^{1}$Faculty of Automation Control, Electronics and Information Technology, Silesian University of Technology, Gliwice, Poland}%
    \thanks{$^{2}$Corresponding author: {\tt\small rafal.potempa@polsl.pl}}
    \thanks{$^{}$© 2025 IEEE. This is the accepted version of a paper published in the 
    Proceedings of the 2025 \emph{11th International Conference on Control, Decision and Information Technologies (CoDIT)}. The final published version is available at IEEE Xplore: \href{https://doi.org/10.1109/CoDIT66093.2025.11321716}{doi.org/10.1109/CoDIT66093.2025.11321716}.}
}

\begin{document}

\maketitle
\thispagestyle{empty}
\pagestyle{empty}

\begin{abstract}

This study aims to introduce the FRQI Pairs method to a wider audience, a novel approach to image classification using Quantum Recurrent Neural Networks (QRNN) with Flexible Representation for Quantum Images (FRQI).

The study highlights an innovative approach to use quantum encoded data for an image classification task, suggesting that such quantum-based approaches could significantly reduce the complexity of quantum algorithms. Comparison of the FRQI Pairs method with contemporary techniques underscores the promise of integrating quantum computing principles with neural network architectures for the development of quantum machine learning.

\end{abstract}
\section{Introduction}

Quantum image processing is a promising field in quantum computing. Starting with Venegas-Andraca and Bose's \cite{venegas2003storing} qubit lattice representation for quantum image encoding, the description of quantum images came right after this \cite{latorre2005image}. Quantum states describe patterns for two main reasons: to improve classification efficiency \cite{wiebe2015quantum} and to provide valuable models for traditional issues \cite{tanaka2007quantum}. The Flexible Representation for Quantum Images (FRQI) was first proposed in \cite{le2011flexible} and was further developed in \cite{le2011flexible_2}. This research focuses on quantum image encoding and quantum machine learning classification methods applied to the MNIST data set.

The field of quantum machine learning is expanding rapidly, and new methods emerge, sometimes inspired by the classical machine learning methods or some developed purely for quantum computers. Some of the recent methods are inspired by traditional machine learning techniques, e.g., quantum state vector machine (QSVM) \cite{delilbasic2021quantum,li2015experimental,rana2022comparative,rebentrost2014quantum}, quantum $k$-nearest neighbors (QKNN) \cite{dang2018image,ruan2017quantum,wang2019improved}
and quantum nearest mean classifier (QNMC) \cite{santucci2017quantum,sergioli2017quantum,sergioli2018quantum}. In the family of deep methods: variational quantum circuits (VQC) \cite{potempa2022comparing,skolik2021layerwise,wang2022development} inspired by classical neural networks, quantum tensor networks (QTN) \cite{grant2018hierarchical,huang2021variational,lazzarin2022multi}, quantum convolutional neural networks (QCNN) \cite{cong2019quantum,huang2022variational,hur2022quantum,shende2004minimal,zheng2023design}, random quantum neural networks (RQNN) \cite{konar2022random} and quantum recurrent neural networks (QRNN) \cite{bausch2020recurrent,potempa2022recognition}. 

Among the last group, there appears to be no agreement on how to phrase the architecture name, namely the words \emph{quantum} and \emph{recurrent} appear in both combinations, resulting in the recurrent models being referred to as QRNN or RQNN. The authors of this paper will use the abbreviation QRNN for recurrent networks and RQNN for random networks.
\section{Subject}

The prototyping results of this paper are based on the thesis \cite{potempa2022recognition}. The work proposes a novel approach to the classification of quantum-encoded images using QRNN \cite{bausch2020recurrent} with the input data encoded in a quantum way, using the FRQI \cite{le2011flexible}. The use of quantum encoding allows for a futuristic assumption that the classification is performed on some universal quantum computer where the encoded data is stored in a quantum memory.

However, such devices are not available in the Noisy Intermediate Scale Quantum (NISQ) \cite{preskill2018quantum} era, and the disposal of the data preprocessing steps required every time an image is loaded into the quantum system would significantly reduce the classical processing overhead.

\subsection{MNIST Database}

The Modified National Institute of Standards and Technology (MNIST) database contains handwritten digits stored in the form of $28 \times 28$ pixel images with $0...255$ values representing the pixel intensity. A scaled-down sample image is shown in \cref{fig:frqi-5-images}. The data set has been chosen due to its wide use for benchmark purposes, hence allowing for broad comparison with classical and quantum methods. The recent criticism of MNIST benchmarks for quantum machine learning algorithms \cite{bowles2024better} suggests that further research is needed to test the approach on different datasets and from other perspectives.

\subsection{Flexible Representation of Quantum Images}\label{sub:FRQI}

The FRQI method proposed in \cite{le2011flexible} allows efficient storage of single channel images using only $\mathcal{O}(\ceil{\log_{2}n})$ qubits, where $n$ is roughly the side length of the image. The method also allows the easy manipulation of the image properties, thus placing it and its descendants as a versatile image storing and manipulation method for a quantum computer. The method has since been improved to support multichannel images in further works \cite{khan2019improved,li2018color,nasr2021efficient}. Since the MNIST database used for classification includes only grayscale images, the FRQI method for encoding has been used for simplicity.

\begin{figure}[t]
    \centering
    \includegraphics[width=0.26\textwidth]{
        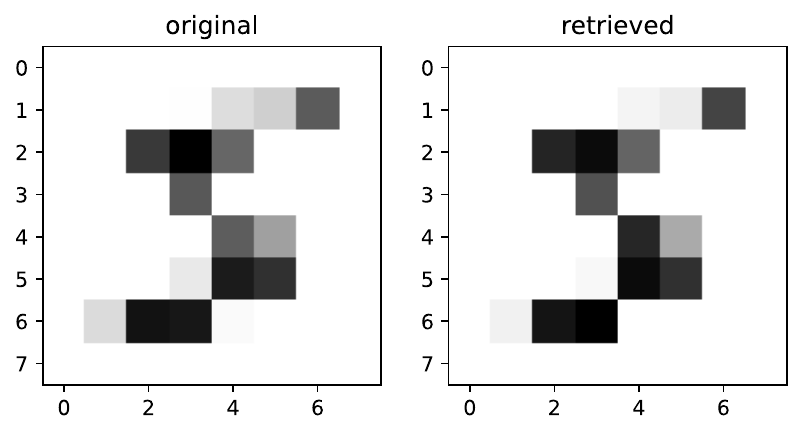}
    \includegraphics[width=0.17\textwidth]{
        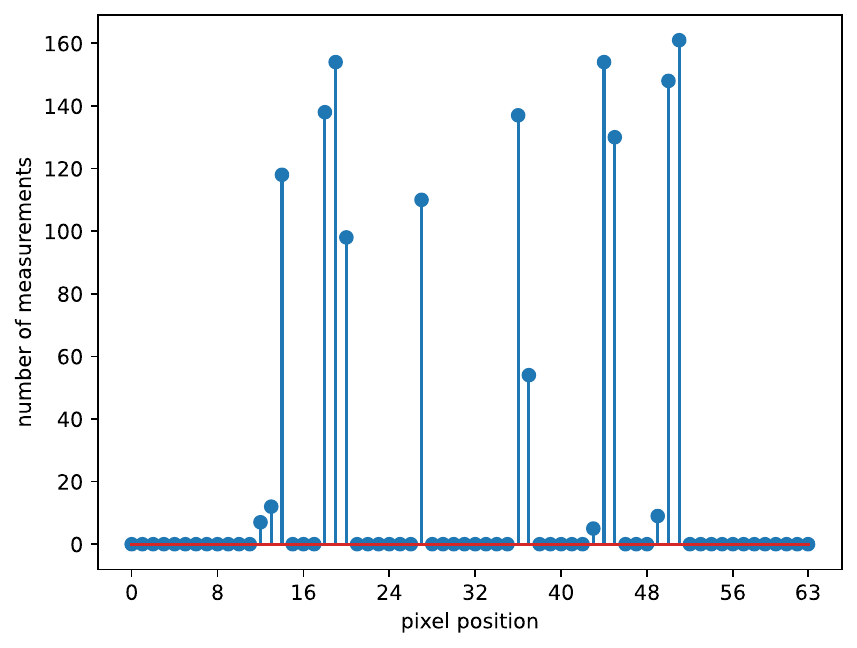}
    \caption{Comparison of the scaled image from MNIST database (left) and its retrieved version from FRQI encoding (center). Aggregated results of $10_000$ measurements of the FRQI representation of the image (right).}
    \label{fig:frqi-5-images}
\end{figure}

The method encodes the position of the image pixel using $\ceil{\log_{2}W}$ and $\ceil{\log_{2}H}$ qubits for the width and height of the image, respectively, plus an additional single qubit for color value encoding $\nu = \ceil{\log_{2}W} + \ceil{\log_{2}H} + 1$.
Assuming that the image is bounded by a square envelope of dimension $2^n$, the resulting number of qubits is
\begin{equation} \label{eqn:frqi-n-qubits}
    \nu = 2n + 1.
\end{equation}
We can describe the resulting FRQI image state $\ket{I}$ as
\begin{equation} \label{eqn:frqi-state}
    \ket{I(\boldsymbol{\theta})} = \frac{1}{2^{n}} \sum_{x=0}^{2^{2n}-1}(\cos\theta_x\ket{0} + \sin\theta_x\ket{1}) \otimes \ket{x},
\end{equation}
where $\theta_x \in [0, \frac{\pi}{2}]$ and $x \in \{ 0, 1, \ldots , 2^{2n}-1\}$. The original MNIST pixel values are integer values from the range $\{0, 1, \ldots 255\}$, thus they have to be uniformly scaled into $[0, \frac{\pi}{2}]$. 
An example of a scaled MNIST image, with its FRQI measurements and retrieved version, is shown in \cref{fig:frqi-5-images}.

\subsection{Quantum Recurrent Neural Networks}

\cite{bausch2020recurrent} presents the first QRNN model capable of performing complex tasks such as sequence learning and digit classification. This QRNN utilizes an enhanced version of a quantum neuron presented in \cite{cao2017quantum} with amplitude amplification to create a nonlinear activation function. The model was tested on various tasks, including memorization, sequence prediction, and MNIST digits dataset classification, demonstrating its capability to handle high-dimensional training data. The QRNN implemented using PyTorch programming library \cite{paszke2019pytorch} shows results that have an impact on quantum machine learning, especially in managing long sequence data without the gradient vanishing problem typical in classical RNNs.

\section{Methods}

\subsection{Examined Models}

The thesis \cite{potempa2022recognition} introduces a quantum-specific approach to combine images encoded in a quantum way with QRNN \cite{bausch2020recurrent}. The novelty of the presented approach is the use of input images encoded by the FRQI method as described in \cref{sub:FRQI}. The data set used for prototyping is the MNIST digits, scaled down to $8 \times 8$ for most cases.

The original work presents three models:
\begin{enumerate}
    \item The single-cell model takes as input all the outputs of the FRQI encoding. Information passes through the cell only once.
    \item Naive model, which is an enhanced version of the single-cell model with multiple repetitions of the cell. The author uses two-cell models in his thesis.
    \item The FRQI Pairs model implements the architecture of a full QRNN, where each cell takes as an input combination of parameters responsible for coding color and position.
\end{enumerate}
The author tested all three models with more than 40 different sets of parameters, such as the number of hidden layers in the cell, the size of the input image (in some cases, $28 \times 28$ images are used), and the optimizer learning rate.

\subsection{FRQI Pairs Model}\label{sub:frqi-pairs}

\begin{figure}[t]
    \centering
    \includegraphics[width=0.45\textwidth]{
        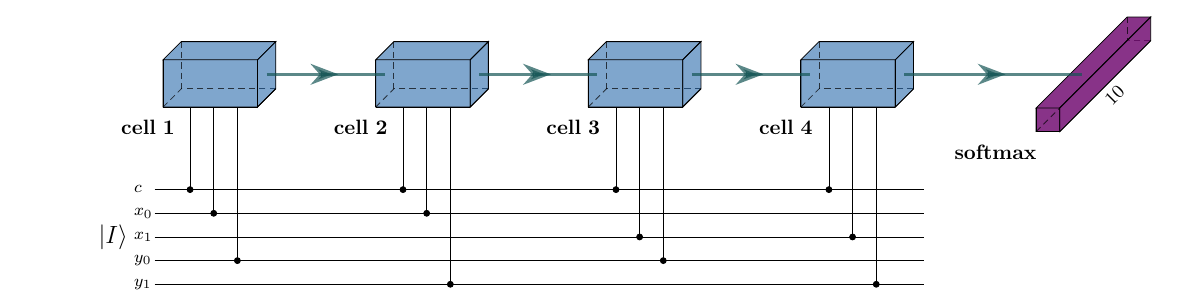}
    \caption{Schematic diagram of the FRQI Pairs Model \cite{potempa2022recognition} for $4 \times 4$ image. The approach uses all combinations of $\{c, x, y\}$ qubits as inputs to QRNN cells, where $x$, $y$ correspond to the X and Y qubits for position encoding.}
    \label{fig:qrnn-frqi}
\end{figure}

The specificity of this approach is the fact that each cell takes as its input two pieces of information (pairs):
\begin{itemize}
    \item the qubits responsible for channel color intensity (e.g., for FRQI -- single qubit for grayscale value),
    \item the qubits responsible for encoding the pixel position (e.g., for 2D images -- two qubits for coding the X-Y position).
\end{itemize}
The total number of cells depends on the number of combinations between $x$ and $y$ qubits, since each cell takes as input one of those combinations. If the image has a dimension of $2^n$ as defined in \cref{eqn:frqi-n-qubits}, the resulting number of cells can be calculated as
\begin{equation} \label{eqn:qrnn-frqi-paris-n-cells}
    N = n^2.
\end{equation}
For comparison, the number of cells in a direct implementation of a QRNN, where each cell takes one pixel of the same image, as in \cite{bausch2020recurrent}, requires an exponentially larger number of cells, compared to \cref{eqn:qrnn-frqi-paris-n-cells}, i.e., $(2^n)^2 = 2^{2n}$. An example diagram of the FRQI Pairs model for a $4 \times 4$ image is shown on \cref{fig:qrnn-frqi}.

\begin{figure}[t]
    \centering
    \includegraphics[width=0.34\textwidth]{
        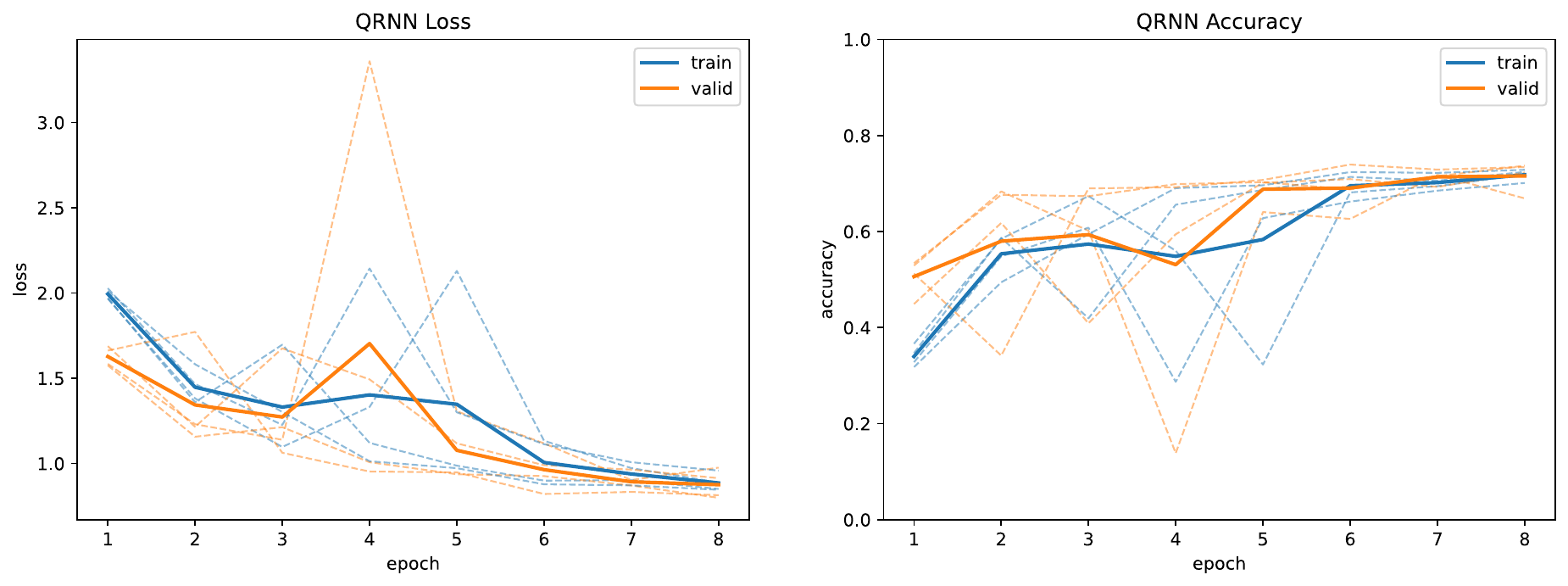}
    \includegraphics[width=0.12\textwidth]{
        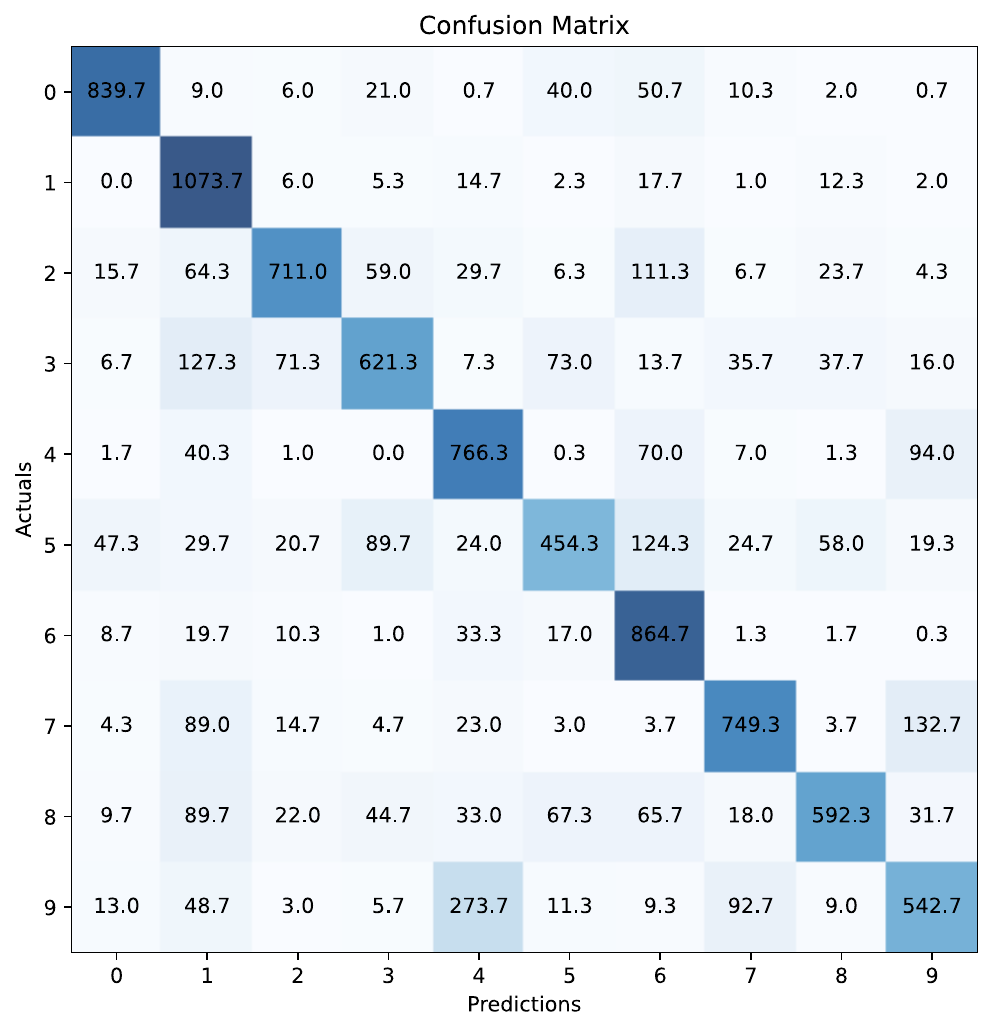}
    \caption{Aggregated results of the four champion FRQI Pairs models training \cite{potempa2022recognition}: loss function - categorical cross-entropy (left), accuracy (center) and average confusion matrix for the test set (right).}
    \label{fig:frqi-pairs-results}
\end{figure}

During the prototyping and tuning phase, increasing the number of QRNN working memory qubits had a positive impact on the test results, with the highest examined number of 4 qubits. Hence, the final QRNN parameters were set to 4 memory qubits and a single deep layer for each cell. The best model used 11 qubits in total (4 QRNN memory + 7 FRQI) and six cells, and its number of trainable parameters was 716 (636 in PQC, 80 for softmax). The model achieved a test accuracy of 74.6\%, which training process and the confusion matrix of the test set is presented in \cref{fig:frqi-pairs-results}.

\section{Comparative analysis}\label{scn:comparative-analysis}

In this section, the authors present the comparison of the results with papers that meet the following criteria:
\begin{enumerate}
    \item the use of the MNIST dataset,
    \item the use of accuracy metrics.
\end{enumerate}
By fulfilling the conditions, it is possible to compare gathered publications with \cite{potempa2022recognition}. The summary of this section is presented in \cref{tab:comparison}.

\begin{table}[t]
    \centering
    \caption{Summary of the works used for comparison in \cref{scn:comparative-analysis}.}
    {\tiny \begin{tabular}{m{0.02\textwidth}|m{0.11\textwidth}|m{0.09\textwidth}|m{0.04\textwidth}|m{0.08\textwidth}}

Ref. & Dataset Modification & Image Encoding & Classifier & Test Accuracy \\
\hline\hline

    \cite{wang2022development} & 
        2000+1000 images\tablefootnote{Training+test set sizes\label{fn:table-train-test-set}}, \newline padded to $32\times 32$, PCA & 
        Amplitude, angle, \newline hybrid encodings & 
        VQDNN & 
        Binary: 99.00\%, \newline 10-class: 80.00\%
    \\ \hline

    \cite{zheng2023design} & 
        1000 images, \newline scaled to $8\times 8$ &
        AQSP & 
        QCNN & 
        Binary: 96.65\% 
    \\ \hline

    \cite{konar2022random} & 
        4-class dataset (24k+4k\footref{fn:table-train-test-set}), noised, classical deep embedding into 128 values & 
        Amplitude encoding & 
        RQNN & 
        4-class: 97.20\%\footref{fn:table-max-accuracy}
    \\ \hline

    \cite{bausch2020recurrent} & 
        Full dataset (60k+10k\footref{fn:table-train-test-set}), \newline scaled to $10 \times 10$, \newline 10-class only: PCA, t-SNE &
        Grayscale value \newline X-gates binary \newline representation & 
        QRNN & 
        Binary: 99.00\%\tablefootnote{Maximum presented test accuracy for given task\label{fn:table-max-accuracy}}, \newline 10-class: 95.00\%\footref{fn:table-max-accuracy}
    \\ \hline

    \cite{potempa2022recognition} & 
        Full dataset (60k+10k\footref{fn:table-train-test-set}), \newline scaled to $8\times 8$ & 
        FRQI & 
        QRNN & 
        10-class: 74.60\%
    \\ \hline

    \cite{siemaszko2023rapid} & 
        800+200 images\footref{fn:table-train-test-set}, \newline 
        scaled to $7\times 7$ &
        Continuous-variable encoding &
        QRNN & 
        Binary: 85.00\% 
    \\ \hline
    
    \cite{wei2022quantum} & 
        Binary: 5000+2100\footref{fn:table-train-test-set}, \newline 
        10-class: full dataset\tablefootnote{Sampled by 100 in each training epoch}, \newline padded up to $32\times 32$ &
        Amplitude encoding & 
        QCNN & 
        Binary: 96.30\%, \newline 10-class: 74.30\%
    \\

\end{tabular}
}
    \label{tab:comparison}
\end{table}

The final results of \cite{potempa2022recognition} show that the proposed model is able to grasp the principles of the underlying data distributions. Its test accuracy is 74.6\%. However, not ideal, its accuracy at the level of magnitude allowing reasonable comparison to three of the presented works, namely \cite{siemaszko2023rapid}, \cite{wang2022development} and \cite{wei2022quantum}.

The work presented in \cite{siemaszko2023rapid} implements a QRNN, which was tested against a binary classification of the MNIST digits $3$ and $6$. The results confirm the QRNN utility for the classification of handwritten digits. However, they suggest that the classical LSTM model with a similar number of parameters performs better, which might suggest that the QRNN model is inefficient in image classification.

\cite{wei2022quantum} represents a QCNN, which is a different family of models, yet the model was trained to classify all ten digits, and except for the 0-padding to $32 \times 32$ size, no initial data transformation was performed. The numbers of trainable parameters of the model from \cite{potempa2022recognition} and \cite{wei2022quantum} are at a similar level: 716 and 379, respectively. For both models, the final accuracy is also similar: 74.6\% and 74.3\%, while the data used to train the final model have a higher resolution in the case of \cite{wei2022quantum}. Both results suggest that the QRNN model might have a higher learning potential than what was achieved in \cite{potempa2022recognition}, and may achieve even better results for full-sized images. From \cref{eqn:frqi-n-qubits} and \cref{eqn:qrnn-frqi-paris-n-cells} we find that the $28 \times 28$ FRQI encoding uses 11 qubits, so the circuit will use 15 or more qubits, depending on the number of memory lanes, while the number of cells will increase to 25. The increase in the number of cells and qubits will result in a higher number of the model's trainable parameters. However, it will also increase the model's training capacity and potentially improve the results.

The authors of \cite{wang2022development} presented a VQDNN model, which resembles a densely connected network. They managed to heavily reduce the number of qubits required to operate the network, providing a solution that uses only ten qubits for 10-class problems. For the features extracted using the PCA, the authors experimented with amplitude and angle encoding, as well as their combination. The total number of parameters for the deepest model was 430. The depth of the model increases the model's training capacity, and the best model achieved approximately 80\% accuracy. The proposed model of the VQDNN network proved its potential for experimentation to replace the internals of the QRNN cell from \cite{potempa2022recognition}.

The use of PCA as the feature extraction method should also be considered in the further development of FRQI-QRNN models, as its impact was shown in \cite{bausch2020recurrent} and \cite{wang2022development}. However, the method might not be directly applicable to the problem \cite{potempa2022recognition} is trying to solve, i.e., the quantum ML model operating on already quantum-encoded data. Despite the fact that the PCA method can help reduce the dimensionality of feature spaces, thus reducing the model complexity and allowing for higher performance, with the same number of trainable parameters--one can imagine that the data stored in future quantum memory would be encoded and compressed using different means. The applicability of the quantum version of PCA \cite{lloyd2014quantum,ostaszewski2015quantum} or other methods such as \cite{li2021resonant} to the feature preprocessing phase should also be explored.

Other works that use the amplitude encoding method for the classification of MNIST datasets are \cite{konar2022random,zheng2023design}. They both present classifier types different from QRNN, but the authors managed to achieve high test accuracy on limited datasets. Random QNN \cite{konar2022random} shows high robustness against noisy data. However, the image embedding is performed by a classical densely connected layer, which, similarly to PCA, offloads some part of the solution to the classical part.

\cite{zheng2023design} proposed the Approximation Quantum State Preparation (AQSP) method for image encoding, which uses a simulated quantum circuit to train the image representation and has a time complexity of only $\mathcal{O}(n)$. The authors also combine the AQSP method with the proposed QCNN framework. The design of a hybrid recurrent network with convolutional and pooling input layers might help make the FRQI-QRNN architecture less dependent on the input image size and reduce the complexity of the recurrent part.

\section{Conclusions}

The presented state-of-the-art methods trained to solve the MNIST classification task prove that the area of Quantum Machine Learning has the potential to solve real-life machine learning problems.

The new FRQI Pairs architecture, presented in \cite{potempa2022recognition}, requires an exponentially lower number of recurrent cells compared to \cite{bausch2020recurrent} ($n^2$ vs. $2^{2n}$ \cref{eqn:qrnn-frqi-paris-n-cells}) which may lead to shorter execution times thus a lower chance of decoherence, and a higher computational efficiency during the inference phase if applied to real-world problems.

An important aspect of the presented architecture is that it utilizes a well-grounded FRQI method as its input, which allows to use the method's data encoding advantage over the classical representation. This saves the processing time at the cost of slightly larger number of qubits, but without the need to leave the quantum realm to use the classical pixel values for qubit encoding. Assuming that in the future we would be able to persistently store data in quantum form, the FRQI Pairs method would be a solid starting point in the development of fully quantum neural networks.

Another important feat of the FRQI Pairs method is that it managed to train on the full MNIST dataset and has proven performance comparable to other methods like \cite{ostaszewski2015quantum,wang2022development}. Many other presented works struggle to capture the idea of MNIST data base, which is to create a dataset for benchmarking machine learning models and ideas based on reliable and repeatable dataset. MNIST is meant to be taken as a whole, to enable direct comparison with numerous classical models.

Although areas were left to improve compared to \cite{bausch2020recurrent}, the method has the potential to be extended by various preprocessing routines \cite{bausch2020recurrent,li2021resonant,lloyd2014quantum,ostaszewski2015quantum} and layer/model architectures \cite{wang2022development,wei2022quantum,zheng2023design}. The approach of the other works to the problem suggests that the research from \cite{potempa2022recognition} should also be extended to a binary image classification case for broader comparison possibilities.

% \input{sections/99_guidelines}

% \addtolength{\textheight}{-12cm}  % This command serves to balance the column lengths

% \input{sections/99_appendix}
\section*{ACKNOWLEDGMENT}

The authors acknowledge that this paper has been written based on the results achieved within the OptiQ project. This project has received funding from the European Union’s Horizon Europe program under grant agreement No 101080374-OptiQ.

Supplementarily, the project is co-financed from the resources of the Polish Ministry of Science and Higher Education in the framework of the International Co-financed Projects program.

Disclaimer. Funded by the European Union. However, views and opinions expressed are those of the author(s) alone and do not necessarily reflect those of the European Union or the European Research Executive Agency (REA – granting authority). Neither the European Union nor the granting authority can be held responsible for them.

\bibliographystyle{ieee}
\bibliography{refs}

@article{bausch2020recurrent,
  title   = {Recurrent quantum neural networks},
  author  = {Bausch, Johannes},
  journal = {Advances in neural information processing systems},
  volume  = {33},
  pages   = {1368--1379},
  year    = {2020}
}

@article{bowles2024better,
  title   = {Better than classical? the subtle art of benchmarking quantum machine learning models},
  author  = {Bowles, Joseph and Ahmed, Shahnawaz and Schuld, Maria},
  journal = {arXiv preprint arXiv:2403.07059},
  year    = {2024}
}

@article{cao2017quantum,
  title   = {Quantum neuron: an elementary building block for machine learning on quantum computers},
  author  = {Cao, Yudong and Guerreschi, Gian Giacomo and Aspuru-Guzik, Al{\'a}n},
  journal = {arXiv preprint arXiv:1711.11240},
  year    = {2017}
}

@article{cong2019quantum,
  title     = {Quantum convolutional neural networks},
  author    = {Cong, Iris and Choi, Soonwon and Lukin, Mikhail D},
  journal   = {Nature Physics},
  volume    = {15},
  number    = {12},
  pages     = {1273--1278},
  year      = {2019},
  publisher = {Nature Publishing Group UK London}
}

@article{dang2018image,
  title     = {Image classification based on quantum K-Nearest-Neighbor algorithm},
  author    = {Dang, Yijie and Jiang, Nan and Hu, Hao and Ji, Zhuoxiao and Zhang, Wenyin},
  journal   = {Quantum Information Processing},
  volume    = {17},
  pages     = {1--18},
  year      = {2018},
  publisher = {Springer}
}

@inproceedings{delilbasic2021quantum,
  title        = {Quantum support vector machine algorithms for remote sensing data classification},
  author       = {Delilbasic, Amer and Cavallaro, Gabriele and Willsch, Madita and Melgani, Farid and Riedel, Morris and Michielsen, Kristel},
  booktitle    = {2021 IEEE International Geoscience and Remote Sensing Symposium IGARSS},
  pages        = {2608--2611},
  year         = {2021},
  organization = {IEEE}
}

@article{grant2018hierarchical,
  title     = {Hierarchical quantum classifiers},
  author    = {Grant, Edward and Benedetti, Marcello and Cao, Shuxiang and Hallam, Andrew and Lockhart, Joshua and Stojevic, Vid and Green, Andrew G and Severini, Simone},
  journal   = {npj Quantum Information},
  volume    = {4},
  number    = {1},
  pages     = {65},
  year      = {2018},
  publisher = {Nature Publishing Group UK London}
}

@article{huang2021variational,
  title     = {Variational quantum tensor networks classifiers},
  author    = {Huang, Rui and Tan, Xiaoqing and Xu, Qingshan},
  journal   = {Neurocomputing},
  volume    = {452},
  pages     = {89--98},
  year      = {2021},
  publisher = {Elsevier}
}

@article{huang2022variational,
  title     = {Variational convolutional neural networks classifiers},
  author    = {Huang, Fangyu and Tan, Xiaoqing and Huang, Rui and Xu, Qingshan},
  journal   = {Physica A: Statistical Mechanics and its Applications},
  volume    = {605},
  pages     = {128067},
  year      = {2022},
  publisher = {Elsevier}
}

@misc{hur2022quantum,
  title     = {Quantum convolutional neural network for classical data classification, Quantum Mach},
  author    = {Hur, T and Kim, L and Park, DK},
  year      = {2022},
  publisher = {Intell}
}

@article{khan2019improved,
  title     = {An improved flexible representation of quantum images},
  author    = {Khan, Rabia Amin},
  journal   = {Quantum Information Processing},
  volume    = {18},
  pages     = {1--19},
  year      = {2019},
  publisher = {Springer}
}

@article{konar2022random,
  title   = {Random quantum neural networks (RQNN) for noisy image recognition},
  author  = {Konar, Debanjan and Gelenbe, Erol and Bhandary, Soham and Sarma, Aditya Das and Cangi, Attila},
  journal = {arXiv preprint arXiv:2203.01764},
  year    = {2022}
}

@article{latorre2005image,
  title   = {Image compression and entanglement},
  author  = {Latorre, Jose I},
  journal = {arXiv preprint quant-ph/0510031},
  year    = {2005}
}

@article{lazzarin2022multi,
  title     = {Multi-class quantum classifiers with tensor network circuits for quantum phase recognition},
  author    = {Lazzarin, Marco and Galli, Davide Emilio and Prati, Enrico},
  journal   = {Physics Letters A},
  volume    = {434},
  pages     = {128056},
  year      = {2022},
  publisher = {Elsevier}
}

@article{le2011flexible,
  title     = {A flexible representation of quantum images for polynomial preparation, image compression, and processing operations},
  author    = {Le, Phuc Q and Dong, Fangyan and Hirota, Kaoru},
  journal   = {Quantum Information Processing},
  volume    = {10},
  number    = {1},
  pages     = {63--84},
  year      = {2011},
  publisher = {Springer}
}

@article{le2011flexible_2,
  title     = {A flexible representation and invertible transformations for images on quantum computers},
  author    = {Le, Phuc Q and Iliyasu, Abdullahi M and Dong, Fangyan and Hirota, Kaoru},
  journal   = {New Advances in Intelligent Signal Processing},
  pages     = {179--202},
  year      = {2011},
  publisher = {Springer}
}

@article{li2015experimental,
  title     = {Experimental realization of a quantum support vector machine},
  author    = {Li, Zhaokai and Liu, Xiaomei and Xu, Nanyang and Du, Jiangfeng},
  journal   = {Physical review letters},
  volume    = {114},
  number    = {14},
  pages     = {140504},
  year      = {2015},
  publisher = {APS}
}

@article{li2018color,
  title     = {Color image representation model and its application based on an improved FRQI},
  author    = {Li, Panchi and Liu, Xiande},
  journal   = {International Journal of Quantum Information},
  volume    = {16},
  number    = {01},
  pages     = {1850005},
  year      = {2018},
  publisher = {World Scientific}
}

@article{li2021resonant,
  title     = {Resonant quantum principal component analysis},
  author    = {Li, Zhaokai and Chai, Zihua and Guo, Yuhang and Ji, Wentao and Wang, Mengqi and Shi, Fazhan and Wang, Ya and Lloyd, Seth and Du, Jiangfeng},
  journal   = {Science Advances},
  volume    = {7},
  number    = {34},
  pages     = {eabg2589},
  year      = {2021},
  publisher = {American Association for the Advancement of Science}
}

@article{lloyd2014quantum,
  title     = {Quantum principal component analysis},
  author    = {Lloyd, Seth and Mohseni, Masoud and Rebentrost, Patrick},
  journal   = {Nature Physics},
  volume    = {10},
  number    = {9},
  pages     = {631--633},
  year      = {2014},
  publisher = {Nature Publishing Group UK London}
}

@article{nasr2021efficient,
  title     = {Efficient representations of digital images on quantum computers},
  author    = {Nasr, Norhan and Younes, Ahmed and Elsayed, Ashraf},
  journal   = {Multimedia Tools and Applications},
  volume    = {80},
  pages     = {34019--34034},
  year      = {2021},
  publisher = {Springer}
}

@article{ostaszewski2015quantum,
  title        = {Quantum image classification using principal component analysis},
  author       = {Ostaszewski, Mateusz and Sadowski, Przemysław and Gawron, Piotr},
  volume       = {vol. 27},
  number       = {No 1},
  journal      = {Theoretical and Applied Informatics},
  pages        = {1-12},
  howpublished = {online},
  year         = {2015},
  publisher    = {Institute of Theoretical and Applied Informatics of Polish Academy of Science}
}

@article{paszke2019pytorch,
  title   = {Pytorch: An imperative style, high-performance deep learning library},
  author  = {Paszke, Adam and Gross, Sam and Massa, Francisco and Lerer, Adam and Bradbury, James and Chanan, Gregory and Killeen, Trevor and Lin, Zeming and Gimelshein, Natalia and Antiga, Luca and others},
  journal = {Advances in neural information processing systems},
  volume  = {32},
  year    = {2019}
}

@inproceedings{potempa2022comparing,
  title        = {Comparing concepts of quantum and classical neural network models for image classification task},
  author       = {Potempa, Rafa{\l} and Porebski, Sebastian},
  booktitle    = {Progress in Image Processing, Pattern Recognition and Communication Systems: Proceedings of the Conference (CORES, IP\&C, ACS)-June 28-30 2021 12},
  pages        = {61--71},
  year         = {2022},
  organization = {Springer}
}

@mastersthesis{potempa2022recognition,
  author = {Potempa, Rafa{\l} and Porebski, Sebastian},
  title  = {Recognition of quantum images using the {FRQI} algorithm and the {QRNN} classifier},
  school = {Silesian University of Technology},
  year   = {2022}
}

@article{preskill2018quantum,
  title     = {Quantum computing in the {NISQ} era and beyond},
  author    = {Preskill, John},
  journal   = {Quantum},
  volume    = {2},
  pages     = {79},
  year      = {2018},
  publisher = {Verein zur F{\"o}rderung des Open Access Publizierens in den Quantenwissenschaften}
}

@article{rana2022comparative,
  title     = {A comparative study of quantum support vector machine algorithm for handwritten recognition with support vector machine algorithm},
  author    = {Rana, Anurag and Vaidya, Pankaj and Gupta, Gaurav},
  journal   = {Materials Today: Proceedings},
  volume    = {56},
  pages     = {2025--2030},
  year      = {2022},
  publisher = {Elsevier}
}

@article{rebentrost2014quantum,
  title     = {Quantum support vector machine for big data classification},
  author    = {Rebentrost, Patrick and Mohseni, Masoud and Lloyd, Seth},
  journal   = {Physical review letters},
  volume    = {113},
  number    = {13},
  pages     = {130503},
  year      = {2014},
  publisher = {APS}
}

@article{ruan2017quantum,
  title     = {Quantum algorithm for k-nearest neighbors classification based on the metric of hamming distance},
  author    = {Ruan, Yue and Xue, Xiling and Liu, Heng and Tan, Jianing and Li, Xi},
  journal   = {International Journal of Theoretical Physics},
  volume    = {56},
  pages     = {3496--3507},
  year      = {2017},
  publisher = {Springer}
}

@article{santucci2017quantum,
  title     = {Quantum minimum distance classifier},
  author    = {Santucci, Enrica},
  journal   = {Entropy},
  volume    = {19},
  number    = {12},
  pages     = {659},
  year      = {2017},
  publisher = {MDPI}
}

@article{sergioli2017quantum,
  title     = {A quantum-inspired version of the classification problem},
  author    = {Sergioli, Giuseppe and Bosyk, Gustavo Martin and Santucci, Enrica and Giuntini, Roberto},
  journal   = {International Journal of Theoretical Physics},
  volume    = {56},
  pages     = {3880--3888},
  year      = {2017},
  publisher = {Springer}
}

@article{sergioli2018quantum,
  title     = {A quantum-inspired version of the nearest mean classifier},
  author    = {Sergioli, Giuseppe and Santucci, Enrica and Didaci, Luca and Miszczak, Jaros{\l}aw A and Giuntini, Roberto},
  journal   = {Soft Computing},
  volume    = {22},
  pages     = {691--705},
  year      = {2018},
  publisher = {Springer}
}

@article{shende2004minimal,
  title     = {Minimal universal two-qubit controlled-NOT-based circuits},
  author    = {Shende, Vivek V and Markov, Igor L and Bullock, Stephen S},
  journal   = {Physical Review A},
  volume    = {69},
  number    = {6},
  pages     = {062321},
  year      = {2004},
  publisher = {APS}
}

@article{siemaszko2023rapid,
  title     = {Rapid training of quantum recurrent neural networks},
  author    = {Siemaszko, Micha{\l} and Buraczewski, Adam and Le Saux, Bertrand and Stobi{\'n}ska, Magdalena},
  journal   = {Quantum Machine Intelligence},
  volume    = {5},
  number    = {2},
  pages     = {31},
  year      = {2023},
  publisher = {Springer}
}

@article{skolik2021layerwise,
  title     = {Layerwise learning for quantum neural networks},
  author    = {Skolik, Andrea and McClean, Jarrod R and Mohseni, Masoud and van der Smagt, Patrick and Leib, Martin},
  journal   = {Quantum Machine Intelligence},
  volume    = {3},
  pages     = {1--11},
  year      = {2021},
  publisher = {Springer}
}

@inproceedings{tanaka2007quantum,
  title     = {A quantum-statistical-mechanical extension of Gaussian mixture model},
  author    = {Tanaka, Kazuyuki and Tsuda, K},
  booktitle = {Proc. Int. Workshop on Statistical-Mechanical Informatics},
  volume    = {95},
  year      = {2007}
}

@inproceedings{venegas2003storing,
  title        = {Storing, processing, and retrieving an image using quantum mechanics},
  author       = {Venegas-Andraca, Salvador E and Bose, Sougato},
  booktitle    = {Quantum information and computation},
  volume       = {5105},
  pages        = {137--147},
  year         = {2003},
  organization = {SPIE}
}

@article{wang2019improved,
  title     = {Improved handwritten digit recognition using quantum k-nearest neighbor algorithm},
  author    = {Wang, Yuxiang and Wang, Ruijin and Li, Dongfen and Adu-Gyamfi, Daniel and Tian, Kaibin and Zhu, Yixin},
  journal   = {International Journal of Theoretical Physics},
  volume    = {58},
  pages     = {2331--2340},
  year      = {2019},
  publisher = {Springer}
}

@article{wang2022development,
  title     = {Development of variational quantum deep neural networks for image recognition},
  author    = {Wang, Yunqian and Wang, Yufeng and Chen, Chao and Jiang, Runcai and Huang, Wei},
  journal   = {Neurocomputing},
  volume    = {501},
  pages     = {566--582},
  year      = {2022},
  publisher = {Elsevier}
}

@article{wei2022quantum,
  title     = {A quantum convolutional neural network on {NISQ} devices},
  author    = {Wei, ShiJie and Chen, YanHu and Zhou, ZengRong and Long, GuiLu},
  journal   = {AAPPS Bulletin},
  volume    = {32},
  pages     = {1--11},
  year      = {2022},
  publisher = {Springer}
}

@article{wiebe2015quantum,
  title   = {Quantum nearest-neighbor algorithms for machine learning},
  author  = {Wiebe, Nathan and Kapoor, Ashish and Svore, Krysta M},
  journal = {Quantum information and computation},
  volume  = {15},
  number  = {3-4},
  pages   = {318--358},
  year    = {2015}
}

@article{zheng2023design,
  title     = {Design of a quantum convolutional neural network on quantum circuits},
  author    = {Zheng, Jin and Gao, Qing and L{\"u}, Jinhu and Ogorza{\l}ek, Maciej and Pan, Yu and L{\"u}, Yanxuan},
  journal   = {Journal of the Franklin Institute},
  volume    = {360},
  number    = {17},
  pages     = {13761--13777},
  year      = {2023},
  publisher = {Elsevier}
}

\end{document}